\documentclass[aps,prd,reprint,10pt]{revtex4-1}
\usepackage{graphicx} 
\usepackage{comment}
\usepackage[table]{xcolor}
\pdfoutput=1
\begin{document}
\title{Bayesian probability for leptonic CP phase with strong CP prior}
\author{Ravi Kuchimanchi}
\email{ravikuchimanchi1@gmail.com}
\begin{abstract}
 In a world devoid of axions, the smallness of the strong CP phase can have implications for leptonic CP violation being probed by neutrino experiments.  For example, if nature adopted an axionless solution to the strong CP problem, the same symmetries that set the strong CP phase to zero at the tree-level, may also set the leptonic $CP$ phases to zero (mod $\pi$).  This \emph{automatically} happens in the left-right symmetric model when it is extended minimally to solve the strong CP problem by imposition of both $P$ and $CP$.  In the Nelson Barr solution the needed symmetries can be assigned so that leptonic CP violation is not generated. In the minimal left-right symmetric model with P (and not CP), where the strong CP problem remains, leptonic CP phases radiatively generate the strong CP phase in one loop, and therefore they may be absent (negligibly small) in large regions of parameter space.  All these results motivate us to consider a Bayesian prior for leptonic Dirac CP phase $\delta_{CP}$ of the PMNS matrix that has delta function like peaks at CP conserving values of $0$ and $\pi$ on top of a uniform distribution. We evaluate the posterior probability distribution for $\delta_{CP}$ using the current global fit to neutrino experiments and find significant enhancements for the probability that $\delta_{CP}$ is at or negligibly close to $\pi$. We also provide useful tables for the posterior probability considering present and future experimental sensitivities.  
\end{abstract}
\maketitle
\section{Introduction} 

If there is a chink in the armor of the Standard Model, it is the existence of neutrino masses and mixing.  The Standard Model itself does not include right handed neutrinos and therefore it had expected that the neutrinos would be massless. %

The Standard Model (with more particles added to account for neutrino masses) is no more favored than some other models. There is no one standard Lagrangian for which there is consensus in the field, that includes neutrino masses. 
 In fact in the left-right symmetric model~\cite{PhysRevD.10.275,PhysRevD.11.566,Senjanovic:1975rk}, the right handed neutrinos are theoretically required as parity partners of the left-handed neutrinos, and thereby this model anticipated neutrino masses and mixing, which have since been discovered.  Can future experimental results distinguish between different theories? 

One of the last remaining fundamental parameters of nature that is currently being sought by neutrino experiments is the Dirac $CP$ violating phase $\delta_{CP}$ of the PMNS matrix in the leptonic sector. We begin this work by noting that the prior expectation on whether there will be detectable leptonic CP violation or not depends on our theoretical bias. This dependency may turn out to be a blessing because being able to distinguish between theories, or between types of parameter spaces of theories, based on ongoing and future experimental results is a great way to make progress in our understanding of the laws of nature. 

Bayes analysis (see for example Reference~\cite{doi:10.1080/00107514.2012.756312}) helps modify prior beliefs to posterior probabilities based on experimental results. For CP violating phases, the prior beliefs have to do with the largeness of the CKM phase ($\delta_{CKM} \sim 1$), the smallness of the strong CP phase ($|\bar{\theta}| < 10^{-10}$),  and the theoretical lens through which we look at the strong CP problem or incorporate its implications.   CP violation is also needed for baryogenesis, but this does not necessarily depend on the leptonic Dirac phase ($\delta_{CP}$).

We distinguish between two approaches to solve the strong CP problem -- 

(i) Axionic solution -- If we assume Peccei-Quinn symmetry~\cite{1977PhRvL..38.1440P} (and therefore axions~\cite{PhysRevLett.40.223,PhysRevLett.40.279}) then the smallness of $\bar{\theta}$ does not have to do with other $CP$ violating phases and does not constrain leptonic CP violation, which we would then expect be present at the tree-level like $\delta_{CKM}$.  

(ii) Axionless solutions -- On the other hand if the strong CP problem is solved by imposing $CP$ along with $P$ (as in the left-right symmetric model with $P \times CP$~\cite{Kuchimanchi:2010xs}) or $CP$ along with another symmetry (as in Nelson-Barr model~\cite{Nelson:1983zb, PhysRevLett.53.329}), then a mechanism that involves vectorlike quarks that can have CP-violating coupling with the usual quarks is needed to generate the $CKM$ phase. As we recall in the next section leptonic CP phases are not generated at the tree level in the minimal version of the strong CP solving left-right symmetric model with $P \times CP$. In the Nelson-Barr model, and in the non-minimal versions of left-right model with $P \times CP$, depending on how the symmetries are assigned, they may be generated or not~\cite{Kuchimanchi:2012te}. Thus CP violation in the leptonic sector may or may not be generated and the prior belief in these cases would assign a reasonable probability both for tree level leptonic CP conservation, and for its violation.  

Based on the above prior beliefs, using Bayesian analysis, we evaluate the posterior probabilities given the global fit~\cite{Esteban:2020cvm} to the current data from NOvA and T2K neutrino experiments~\cite{Kolupaeva:2020pug,Abe:2019vii} for leptonic $\delta_{CP}$. Our key result is that the probabilities for tree-level leptonic CP conservation are enhanced significantly by the current experimental data, if there are no axions.     We also evaluate the posterior probabilities considering sensitivities of future experiments and find that depending on the results, the enhancement can become even more significant and reach levels above $90\%$, or could diminish to become negligibly small.

We also consider the prior expectation for the case where nature does not adopt a strong CP solution.   We see that the prior for  the minimal left-right symmetric model (with $P$) is different than that for the standard model (with 3 right handed neutrinos added), since in the minimal left-right model the leptonic CP phases can generate the strong CP phase in just one loop and therefore have to be negligibly small in the significant regions of its parameter space.

The rest of the paper is organized as follows.  In Section~\ref{sec:bay} we introduce the Bayesian analysis and discuss the priors based on differing theoretical models.   The prior for strong CP solutions without axions is discussed in Sub-Section~\ref{sec:noaxion} while the case of minimal left-right symmetric model is discussed in Sub-Section~\ref{sec:LR}, and SM and GUTs in Sub-Section~\ref{subsec:SM3}.  In Sectons~\ref{sec:postgauss} and~\ref{sec:postglobal} we discuss the posterior probabilities using Gaussian fits considering future experimental sensitivities and using the current experimental global fit respectively. Section~\ref{sec:utility} discusses the utility of this analysis and finally in Section~\ref{sec:conc}  we summarize our conclusions.

\section{Bayesian prior for $\delta_{CP}$}
\label{sec:bay}

Through experiments we seek to find the unknown parameters of nature.  Through experimental fits we determine which values of theoretical parameters are more likely to give the observed experimental outcomes or data. Through Bayes analysis we can turn this question around and find what we really want to know -- 
that is, which values of parameters are more likely (posterior probability distribution), given the experimental outcomes. 

 In this approach (see for example Ref~\cite{doi:10.1080/00107514.2012.756312} and the section on Mathematical tools, Statistics review in Ref~\cite{Zyla:2020zbs}) we start with a prior probability density for the value of the parameters (in this work, leptonic CP phase $\delta_{CP}$), that can be based on theoretical insight and/or prior experience, and find the posterior probability density given data $p(\delta_{CP}|data)$ (we will use the shorthand notation $p(\delta_{CP})$ for this) as we get more knowledge through the experimentally determined likelihood of data, $L(data|\delta_{CP})$ (we will use the notation $L(\delta_{CP})$ for this).

\begin{equation}
\label{eq:posteq}
p(\delta_{CP}) = L(\delta_{CP}) \times \pi(\delta_{CP}) / \bar{N} 
\end{equation}  
 
where $\pi(\delta_{CP})$ is the prior probability density distribution and $\bar{N}$ is a normalization constant determined so that $\int_0^{2\pi}{p(\delta_{CP}) d\delta_{CP}} = 1.$

The prior has importance. With the same experimental information, depending on the choice of the prior, we may get a deeper insight on the value of parameters.   

What would be an appropriate prior for the CP violating leptonic Dirac phase of the PMNS matrix?  

Let us do a thought experiment.  Consider the situation where prior experiments have discovered that CKM phase and the strong CP phase, both have order 1 values.  Contrast this with another hypothetical situation where CP appears to be conserved in nature and both these phases have been discovered to be consistent with $0$ or $\pi$ to a very high level of accuracy (say to within a hundredth of a degree).  What would be our prior expectation for the leptonic CP phase in both these hypothetical situations?  Will it be the same?

In the first case, nature does not seem to care about CP conservation.  Therefore a uniform distribution for the prior that shows no preference for any particular value of the leptonic CP phase, would be highly motivated.  

On the other hand, for the second situation, if we adopt the same uniform prior, it would mean that we are not  using the knowledge that so far no CP violation has been detected.  In fact there would be leading theories that assume  CP conservation and would predict the absence of leptonic CP violation. Such a prediction would imply a prior with delta function like peaks at the CP conserving values of 0 and 180 degrees. 
However baryon asymmetry in the universe would point to the possibility that CP may be violated. 
 This would motivate a prior that not only has delta function like peaks at 0 and 180 degress, but also has some percentage of the probability that lies in a  uniform distribution: 

\begin{equation}
\label{eq:prioreq}
\pi(\delta_{CP}) = P^{pr}_0 \delta(\delta_{CP}) + P^{pr}_{180} \delta(\delta_{CP}-\pi) + {{1-(P^{pr}_0+P^{pr}_{180})}\over{2\pi}}
 \end{equation}
where $\delta$ is the Dirac delta function and $P^{pr}_0, P^{pr}_{180} \geq 0$ are the prior probabilities of finding $\delta_{CP}$ to be at (or negligibly close to) $0$ and $\pi$ respectively while the last term is the contribution of the uniform distribution. Note that $\int_0^{2\pi}{\pi(\delta_{CP}) d\delta_{CP}} = 1$  and $P^{pr}_{0}+ P^{pr}_{180} \leq 1$. There can be alternate choices for the last term such as a uniform distribution over $sin(\delta_{CP})$ instead of over $\delta_{CP}$, however in this work we stick with the choice made above without loss of generality.

Note that in this work we will be using the delta function as a good approximation for a sharply peaked function of width of a hundredth of a degree or less.  Since the leptonic CP phase is going to be measured only to an accuracy of a few degrees in the foreseeable future, the width of a hundredth of a degree is indeed very much smaller in comparison and therefore the delta function is a very good approximation.

\begin{table}[ht]
	\centering
		\begin{tabular}{p{4cm}p{0.5cm}p{4cm}}
	\hline \hline
		\textbf{Case}  & & \textbf{Prior for leptonic $\mathbf{\delta_{CP}}$} \\
		\hline \\
		
		1)  If \bfseries $\delta_{CKM} \sim \bar{\theta} \sim 1$ \normalfont & \ \  & Uniform distribution with $P^{pr}_0=P^{pr}_{180}=0$\\ \\ \hline 
		 &  & \\
		2)  If $\delta_{CKM} \approx \bar{\theta} \approx 0$ to within a fraction of a degree & \ \ & Mostly expect CP conservation. $P^{pr}_0=P^{pr}_{180}= 0.25~to~ 0.4$ \\ \\ \hline 
		 & & \\
		3)   $\delta_{CKM} \sim 1, |\bar{\theta}| \leq 10^{-10}$ & & \\
		 & & \\
		AXIONS & & \\
		 & & \\
		  If there are axions (Peccei-Quinn symmetry) & & $P^{pr}_0=P^{pr}_{180}=0$ \\
		 & & \\
		  NO AXIONS & &  \\
			& & \\
		  Left Right Symmetric model with $P \times CP$ (see section~\ref{sec:SCPLR}) & & $P^{pr}_0+P^{pr}_{180} \approx 1; P^{pr}_0,P^{pr}_{180}\sim 0.5 $ for minimal model. $P^{pr}_0+P^{pr}_{180}=0.5$ for next to minimal model.  $P^{pr}_0, P^{pr}_{180}\sim 0.05~to~0.25$ (for non-minimal models).\\ 
			& & \\
		  Nelson-Barr completion of SM plus three right handed neutrinos (see section~\ref{sec:NB}) & & $P^{pr}_0 + P^{pr}_{180} = 0.5$ for minimal model.  $P^{pr}_0, P^{pr}_{180}\sim 0.05~to~0.25$ to include non-minimal models. \\
		 & & \\
		  Minimal Left Right Symmetric model with $P$ (see section~\ref{sec:LR}) & & $P^{pr}_0,P^{pr}_{180} \sim 0.25$ \\
		 & & \\

	   SM with right handed neutrinos and GUTs such as $SU(5), SO(10)$ (see section~\ref{subsec:SM3}) & & $P^{pr}_0=P^{pr}_{180}=0$ \\
		 & & \\
				\hline \hline
		\end{tabular}
		\caption{Prior for $\delta_{CP}$ with representative values for $P^{pr}_0, P^{pr}_{180}$ to be used in equation~\ref{eq:prioreq}.}
				\label{tab:prior}

\end{table}	

For the first case of both CKM phase and strong CP phase being order $\sim 1$, we would set $P^{pr}_0=P^{pr}_{180}=0$.   On the other hand if both the CKM and strong CP phase were found to be negligibly close to $0$ or $\pi$, we would use a reasonably high value for $P^{pr}_0+P^{pr}_{180}$ such as $P^{pr}_0=P^{pr}_{180}=0.4$ or we could balance the probabilities so that they are equally shared between the two delta functions and the uniform distribution such as $P^{pr}_0=P^{pr}_{180}=0.25$. What we would use would depend on how we balance the expectation of $CP$ conservation hinted by the absence of CKM and strong CP phase, with the possibility for CP violation that may be needed for generating the baryon asymmetry. As it turns out when we calculate the posterior probabilities, as more and more experiments get done, the terms corresponding to the delta functions get strengthened or weakened, and it doesnt make so much of a difference what exact values we use, as long as they are qualitatively similar. The initial rows of Table~\ref{tab:prior} summarize these hypothetical cases. 

Now we come to the most interesting case of our universe where CP violation has been discovered in the weak interactions, while CP appears to be conserved by the strong interactions to a very high degree. That is the case of $\delta_{CKM} \sim 1$, along with $|\bar{\theta}| \lesssim 10^{-10}$. What does the smallness of the strong CP phase imply?

One of the most popular theories to understand the smallness is the Peccei-Quinn symmetry which requires axions.  If these are discovered, then the smallness of the strong CP phase would not have anything to do with other CP phases in nature, including the leptonic CP phase for which there would be then be no \it a priori \normalfont preferred values, and we would set $P^{pr}_0=P^{pr}_{180}= 0$ (see Table~\ref{tab:prior}). 

However despite a lot of effort, axions have not yet been discovered in nature and those at the scale originally proposed have been ruled out. While axions that are more hidden and therefore eluding discovery are possible, their existence is by no means guaranteed.

In fact the standard model does not include axions, nor do minimal versions of higher theories such as left-right symmetric model or grand unified theories. There are also axionless solutions to the strong CP problem.   

We now discuss the priors for theories that do not include axions. We begin with axionless solutions to the strong CP Problem. 

The priors for theories with no axions are summarized in Table~\ref{tab:prior} for readers who wish to skip directly to Sections~\ref{sec:postgauss} and \ref{sec:postglobal} for the computation of posterior probability distribution using Bayesian analysis.  
  
\subsection{Prior for solutions to the strong CP problem without axions}
\label{sec:noaxion}
\subsubsection{Left Right Symmetric Model with $P \times CP$}
\label{sec:SCPLR}
One of the most beautiful ideas beyond the standard model, is the minimal left-right symmetric model based on $SU(2)_L \times SU(2)_R \times U(1)_{B-L} \times P$ that restores parity as a good symmetry of the Lagrangian,  broken spontaneously. 

The strong CP Phase is odd under parity and therefore is absent in the high energy Lagrangian.  However it is spontaneously generated from other CP phases, when $P$ is spontaneously broken. 

Relatively recently it was noticed that if the minimal left-right symmetric model is extended to also include $CP$ as a symmetry of the Lagrangian (so that we have $SU(2)_L \times SU(2)_R \times U(1)_{B-L} \times P \times CP$)~\cite{Kuchimanchi:2010xs}, then the other phases will be absent and the strong CP phase will not be generated on symmetry breaking, and the strong CP problem is solved without requiring an axion.  However to generate the CKM phase one set of heavy vector like quarks are required, along with a $P$-even $CP$-odd real scalar singlet that couples the heavy quarks to the known quarks.  The CKM phase is generated in the quark sector when the singlet picks up a CP breaking VEV and the heavy quarks are integrated out below this scale. Since the singlet VEV is $P$ symmetric, the quark mass matrices are Hermitian at the tree-level and this generates the CKM phase without generating the strong CP phase, thereby solving the strong CP problem. 

In the minimal version of this model since there is no vector like lepton family that is introduced, the leptons do not couple to the singlet VEV and therefore leptonic CP violation is absent at the tree-level~\cite{Kuchimanchi:2010xs},~\cite{Kuchimanchi:2012te}. Radiative corrections from the quark sector in higher loops, can induce some small amount of leptonic CP violation but this is negligibly small ($\leq 0.01^o$).   

Thus the minimal strong CP solving left-right model with $P$ and $CP$ predicts an absence of measurable leptonic CP phases and we would expect $\delta_{CP} = 0~or~\pi$ to $0.01^o$ or less. Since experiments in the next couple of decades are only likely to measure the leptonic Dirac CP phase to about 3-5 degrees or so, we effectively have a Bayesian prior for $\delta_{CP}$ with delta function like peaks at $0$ and $\pi$, and we can use equation~(\ref{eq:prioreq}) with $P^{pr}_0+P^{pr}_{180} \approx 1$ (say $P^{pr}_0=P^{pr}_{180}\approx 0.5$) for this minimal model. Note that in this work the delta function in equation~(\ref{eq:prioreq}) is understood to be a  very good approximation for a sharply peaked function with a width of less than a tenth of a degree, which is a negligibly small width.  

We can go beyond the minimal version, and hope to introduce CP violation in the leptonic sector by adding a vector-like leptonic family, analogous to the vector-like quark family.  However there is a caveat.  Because of the presence of both Majorana and Dirac Yukawa couplings, the new vector-like leptonic family can either be chosen to have an intrinsic parity phase ($\eta_{L'}$) of $1$ or $i$ as shown in Ref~\cite{Kuchimanchi:2012te}, so that under $P$, $L'_{L,R} \rightarrow L'_{R,L}$ or $L'_{L,R} \rightarrow i L'_{R,L}$ respectively, where $L'$ is the heavy lepton.     We do not know a priori what the assignment of the intrinsic phase is, and we would consider both possibilities equally as a prior.  In the first case the vector like lepton with real intrinsic parity ($\eta_{L'}=1$), has a Yukawa coupling with the CP odd scalar singlet and existing leptons, which also have real intrinsic parities, and therefore generates the CP phase.  However if the intrinsic parity of the heavy lepton is $i$, its coupling with usual leptons with real intrinsic parities is forbidden by $P^2$ operator which is $\eta_{L'}^2 = -1$ for the heavy leptons while is $1$ for all other particles. $P^2$ is thus a $Z_2$ symmetry that is automatic in this model (as $P$ implies $P^2$), and it can be identified with the $Z_2$ symmetry that produces a dark sector or dark matter. Thus if $\eta_{L'} = i$, the heavy leptons belong in the dark sector or form dark matter, do not have yukawa couplings with the usual leptons, and therefore no leptonic CP violation is introduced!  

Thus we have a choice of intrinsic parities to provide to the heavy leptons which is not there with the quarks.  Depending on the choice, we either have leptonic CP violation (for $\eta_{L'}= 1$) or we have tree-level leptonic CP conservation (for $\eta_{L'} = i$). Note that due to the absence of Majorana type Yukawa couplings in the quark sector, there is a $U(1)_B$ global symmetry acting on quarks alone, that can rotate the imaginary intrinsic parity phase away in the quark sector -- therefore this choice is not available in the quark sector~\cite{Kuchimanchi:2012te}.

Thus for the Bayesian prior for leptonic $\delta_{CP}$ we would expect equal probability for leptonic CP violation and tree-level conservation. Thus  $P^{pr}_0+P^{pr}_{180}=0.5$ (or $P^{pr}_0=P^{pr}_{180}=0.25$) is highly motivated for this model.   

If we wish to be more conservative, we can consider more non-minimal models, such as adding more vectorlike lepton families and again choosing intrinsic parities $1$ or $i$ for them.  More vectorlike lepton families would reduce the prior probability for CP conservation, and we could consider a lower value of $P^{pr}_0=P^{pr}_{180}=0.05$,  but anything much lower would be hard to justify as a prior for this model.  

Considering the above, in Table~\ref{tab:prior} we have used the range $P^{pr}_0, P^{pr}_{180} = 0.05~to~0.25$ for the strong CP solving left-right model with $P$ and $CP$, while noting for the minimal version of this model (with no vector like lepton family) CP conservation is automatic ($P^{pr}_0+P^{pr}_{180}=1$).  

\subsubsection{Nelson-Barr model}
\label{sec:NB}
Another way to solve the strong CP problem without axions is the Nelson Barr mechanism~\cite{Nelson:1983zb, PhysRevLett.53.329} where $CP$ along with a second global symmetry is introduced.  When $CP$ is spontaneously broken the CKM phase is generated, while the global  symmetry ensures that the quark mass matrix has the Nelson-Barr form with zero entries that mutiply the complex phases when its determinant is taken.  Thus the strong CP phase is not generated at the tree-level.    

The model requires the introduction of a vectorlike heavy quark that couples to the usual quarks via a scalar singlet. When the singlet picks up a CP violating VEV at the spontaneous CP breaking scale, and the vectorlike quarks are integrated out, the CKM phase is generated in the usual quark sector, while there is no strong CP phase generated at the tree level.   

Since the standard model does not have neutrino masses built in its renormalizable Lagrangian, we can consider Nelson-Barr models using the SM group with the addition of 3 right handed neutrinos.  Whether the scalar singlet couples to the neutrinos or not depends on the transformation properties assigned to the leptons under the second global symmetry. 

When the minimal Nelson-Barr model with neutrinos was introduced by Branco~\cite{Branco:2003rt}, the singlet was chosen to have Yukawa couplings with the leptonic sector as well as the quark sector and so the leptonic CP phase was also generated along with the CKM phase.

Inspired by the analysis in the strong CP solving left-right model with $P \times CP$,  it was noted in Ref~\cite{Kuchimanchi:2012te} that even in the Nelson-Barr model,  the required symmetries can  be assigned such that the scalar singlet does not couple to the leptons, and thereby only the CKM phase would be generated and the strong CP and leptonic CP phases both would be absent at the tree-level.   

Since either choice for the singlet (to couple to the leptons or not) is available, apriori there is an equal chance of CP violation as there is of CP conservation in this minimal version of Nelson-Barr model. Thus $P^{pr}_0=P^{pr}_{180} = 0.25$ in equation~(\ref{eq:prioreq}) is highly motivated for the prior in Nelson-Barr model.  In Table~\ref{tab:prior} a more conservative range of $0.05-0.25$ has been mentioned for the prior, to also take into account non-minimal models, such as with additional singlets or heavy leptons.
  
\subsection{Prior for theories without a strong CP solution}
\label{sec:noSCP}

\subsubsection{Minimal Left Rigt symmetric model with $P$}
\label{sec:LR}
We now turn to the minimal left right symmetric model based on $SU(3)_c \times SU(2)_L \times SU(2)_R \times U(1)_{B-L} \times P$.  In this case $\theta_{QCD}$ is absent due to $P$ and $\bar{\theta} = Arg Det M_U M_D$ is generated on spontaneous parity breaking, where $M_{U,D}$ are the up and down sector quark mass matrices. The mass matrices depend on the VEVs of the Higgs fields and in this minimal model there is a $SU(2)_R$ triplet Higgs $\Delta_R$ whose VEV breaks $SU(2)_R$ (and a corresponding $\Delta_L$)  and a bi-doublet $\phi$ of $SU(2)_L \times SU(2)_R$ which contains the standard model Higgs doublet.

The quartic interaction term $\alpha_2 Tr (\Delta_R^\dagger \Delta_R \tilde{\phi}^\dagger\phi)$ is in general CP violating, and $\bar{\theta} \sim \alpha_{2I}$ is generated at the tree level itself~\cite{Kuchimanchi:1995rp,Mohapatra:1995xd}, when $\Delta_R$ and $\phi$ pick up VEVs, where  $\alpha_{2I}$ is the imaginary part of $\alpha_2$, and $\tilde{\phi} = \tau_2 \phi^\star \tau_2$. 

The smallness of $\bar{\theta}$ implies smallness of $\alpha_{2I}$ and this means that not only the strong CP phase, but also the CP violation in Higgs quartic terms must be suppressed in the minimal left-right symmetric model. This would then be an exciting way to test the minimal left-right model, except that the parity breaking or right handed $B-L$ scale could be very high $\lesssim 10^{15}GeV$ or even the Planck scale, and the left-right symmetric Higgs sector may not be accessible to colliders.

However quantum effects can probe higher energies, and recently it was found in~\cite{Kuchimanchi_2015} that $\alpha_{2I}$ and hence $\bar{\theta}$ is generated at the one loop level from the Yukawa couplings in the leptonic sector that provide Majorana masses to the right handed neutrino and Dirac masses to the leptons. Since Yukawa couplings and $\bar{\theta}$ are all dimensionless, the radiative corrections are not suppressed by large mass scales which only appear logarithmically.  So what is really powerful about this result is that regardless of how high the parity breaking or see saw scale is,  there is an unsuppressed one loop correction to $\bar{\theta}$ from CP violating phases in the leptonic Yukawa sector.  

As shown in~\cite{Kuchimanchi_2015} this implies that in significant regions of parameter space of the minimal left-right model the leptonic Dirac CP phase of PMNS matrix $\delta_{CP}$ must be negligibly small, or else it will generate a large strong CP phase in one loop.  Thus we would expect CP conservation in leptonic sector in the most significant parts of the parameter space, and we will take $P^{pr}_{180}\sim P^{pr}_0 \sim 0.25$ in eqn~\ref{eq:prioreq} so that we assign at least a 50\% prior chance for leptonic CP conservation in the minimal left-right model because of the above result (see Table~\ref{tab:prior}).  

It is also possible that not only the imaginary parts, but also the real parts of some of the leptonic Yukawa couplings are both small, in which case there can be CP violating phases, in that part of the parameter space for which we also assign 50\% prior chance.
       
\subsubsection{Standard Model with $\nu_R$ and GUTs}
\label{subsec:SM3}

For completeness we also consider the Standard Model augmented by 3 right handed neutrinos. If SM is viewed only as a low energy effective theory, then if it is completed in the UV by say the left-right model or one of the non-axionic solutions to the strong CP problem, then the priors in the previous section will also apply to the low energy SM.  Therefore in this section,  we will assume the SM is not just a low energy effective theory but also captures the high energy neutrino physics (due to the added $\nu_R$) and is a theory that describes nature all the way up to the Planck scale.

Since there is only the standard model Higgs, there is no CP violation in the Higgs potential, and unlike in the left-right model discussed in the previous subsection, the leptonic CP phases do not induce a $\bar{\theta}$ at the one loop or even two loop level.  Thus leptonic CP violation is not constrained by the strong CP phase, and can be anything.  Therefore we would expect $P^{pr}_{0,180} = 0$ for the SM augmented with 3 $\nu_R$.

Same would be the case if we were to consider a grand unified theory such as $SU(5)$ or $SO(10)$. In such theories we would expect to find leptonic CP phases, just like the CKM phase, and therefore $P^{pr}_{0,180} = 0$.


\section{Posterior probability with Gaussian Likelihoods}
\label{sec:postgauss}

Armed with the priors from the previous section (summarized in Table~\ref{tab:prior}) we can evaluate the posterior probability for $\delta_{CP}$ using equation~(\ref{eq:posteq}), once we have the information from experimental data that is expressed as the likelihood function $L(\delta_{CP})$.  

The experimental data of several neutrino experiments such as $T2K, NOvA$ and future ones such as $DUNE$ and $Hyper K$ is (or will be) often combined using global fits. It is the normal practice to convey the best fit point $\mu$ and standard deviation $\sigma$ of the distribution which can be thought of as approximately being a Gaussian.  

Thus in this section we consider Gaussian likelihoods~\cite{Cowan:2010bz}, and evaluate the posterior probability for different outcomes of $\mu$ and $\sigma$, with a view that it will be useful both for current and future experiments and global fits.  In the next section we use the experimentally determined $\chi^2$ by the Global fit of the current neutrino experimental data from $T2k, NOvA$ and past neutrino experiments to evaluate the posterior.  

Note that in this work we are limiting ourselves to one parameter fits for $\delta_{CP}$.   Our results indicate that it may be interesting for experimental groups to also do multi-parameter analysis using a Bayesian prior for $\delta_{CP}$ along the lines that we consider in this work.

Thus we begin by considering experimental likelihood function for a Gaussian distribution:
\begin{equation}
\label{eq:likely}
L(\delta_{CP}) = c \times g(\delta_{CP}; \mu, \sigma^2) 
\end{equation}
 with $c$ being an arbitrary constant since likehood functions do not need to be normalized to 1, and 
\begin{equation}
\label{eq:gauss}
g(\delta_{CP};\mu,\sigma^2) = (1/\sqrt{2\pi\sigma})e^{-{{1}\over {2}}((\delta_{CP}-\mu) / \sigma)^2} + O(\epsilon)
\end{equation}

In the above $\delta_{CP}, \mu$ and $\sigma$ are expressed in radians and $\delta_{CP}$ ranges from $\mu-\pi$ to $\mu+\pi$.

Since $\delta_{CP}$ is a cyclic parameter, $g$ is technically a wrapped normal distribution and has been written as an ordinary Gaussian plus $O(\epsilon)$.  Since experimental sensitivity for $\delta_{CP}$ has already reached $\sigma \leq 35^o$ or so,   the gaussian and the wrapped normal are almost equal for the $\sigma$ we will be considering, and the difference between them is negligibly small of the $O(\epsilon)$ as indicated in the above equation.  We note for the wrapped normal distribution 
\begin{equation}
\label{eq:norm}
\int^{\mu+\pi}_{\mu-\pi} g(\delta_{CP};\mu,\sigma^2) d\delta_{CP} = 1
\end{equation}

Thus if the experimental best fit is represented by a Gaussian with a mean $\mu$ and standard deviation $\sigma$ we can use the likelihood function in eqn~(\ref{eq:likely}) with~(\ref{eq:gauss}), and the prior distribution given by eqn.~(\ref{eq:prioreq}), and use eqn~(\ref{eq:posteq}) to obtain the posterior probability distribution

\begin{eqnarray}
p(\delta_{CP}) & = & [P^{pr}_0 g(0;\mu,\sigma^2) \delta(\delta_{CP}) + P^{pr}_{180} g(\pi;\mu,\sigma^2) \delta(\delta_{CP}-\pi) \nonumber \\ & + & {{1-(P^{pr}_0+P^{pr}_{180})}\over{2\pi}}g(\delta_{CP};\mu,\sigma^2) ] / N
\end{eqnarray}
where the normalization
\begin{equation}
\label{eq:normN}
N = P^{pr}_0 g(0;\mu,\sigma^2) +  P^{pr}_{180} g(\pi;\mu,\sigma^2) + {{1-(P^{pr}_0+P^{pr}_{180})}\over{2\pi}}
\end{equation}
Note that we have used eqn~(\ref{eq:norm}) and $\int^{\mu+\pi}_{\mu-\pi} p(\delta_{CP}) d\delta_{CP} = 1$, and the constant $c$ has been absorbed in $N$.  .

Using the above two equations and with a little jugglery we can rewrite the above in the useful form
\begin{eqnarray}
\label{eq:post} 
p(\delta_{CP}) & = & P_0 \delta(\delta_{CP}) + P_{180} \delta(\delta_{CP}-\pi) \nonumber \\  & \  & + \left(1-(P_0+P_{180})\right)g(\delta_{CP};\mu,\sigma^2)
 \end{eqnarray}
with 
\begin{equation}
\label{eq:p180}
P_{180} = P^{pr}_{180} g(\pi;\mu,\sigma^2)/N, ~~~P_{0} = P^{pr}_{0} g(0;\mu,\sigma^2)/N
\end{equation}
where $N$ is given by eqn~(\ref{eq:normN}) and we can ignore terms of the order $\epsilon$ in eqn~(\ref{eq:gauss}) for $g$.

$P_0$ and $P_{180}$ are basically the posterior probabilities for $\delta_{CP}$ to be at (or within less than $0.01^o$) of $0^o$ and $180^o$ respectively and are determined by the prior probabilities and the gaussian distribution corresponding to the experimental fit. 


\begin{widetext}{\begin{center}
\begin{table}
\begin{tabular}{ l | l | l | l | l | l | l | l | l | l | l | l | l | l | l | l  }
\hline \hline
$\sigma$   \ \ \     $\mu$ & $180^o$ & $185^o$ & $190^o$ & $195^o$ & $200^o$ & $205^o$ & $210^o$ & $215^o$ & $220^o$ & $225^o$ & $230^o$ & $235^o$ & $240^o$ & $245^o$ & $250^o$ \\ \hline \hline
	$35^o$ & 19\% & 18\% & 18\% & 17\% & 16\% & 15\% & 14\% & 12\% & 11\% & 9\% & 8\% & 6\% & 5\% & 4\% & 3\% \\ \hline
	$30^o$ & 21\% & 21\% & 20\% & 19\% & 18\% & 16\% & 14\% & 12 \% & 10\% & 8\% & 6\% & 5\% & 3\% & 2\% & 2\% \\ \hline
	$25^o$ & 24\% & 24\% & 23\% & 21\% & \cellcolor[HTML]{C0C0C0}19\% & 16\% & 13\% & 11\% & 8\% & 6\% & 4\% & 3\% & 2\% & 1\% & .6\% \\ \hline
	$20^o$ & 29\% & 28\% & 26\% & 23\% & 19\% & 15\% & 11\% & 8\% & 5\% & 3\% & 2\% & 0.9\% & .4\% & 0.2\% & 0.1\% \\ \hline
	$15^o$ & 35\% & 33\% & 30\% & 24\% & 18\% & 12\% & 7\% & 3\% & 1.5\% & 0.6\% & 0.2\% & 0.1\% & 0 & 0 & 0 \\ \hline
	$10^o$ & 44\% & 41\% & 33\% & 21\% & 10\% & 3\% & 1\% & .2\% & 0 & 0  & 0  & 0 & 0 & 0 & 0 \\ \hline
	$5^o$ & 61\% & 49\% & 18\% & 2\% & 0 & 0 & 0 & 0 & 0 & 0 & 0 & 0 & 0 & 0 & 0 \\ \hline 
	$2.5^o$ & 76\% & 30\% & .1\% & 0\% & 0 & 0 & 0 & 0 & 0 & 0 & 0 & 0 & 0 & 0 & 0 \\ \hline \hline
\ \ \ \ \ 	$\mu$ & $180^o$ & $175^o$ & $170^o$ & $165^o$ & $160^o$ & $155^o$ & $150^o$ & $145^o$ & $140^o$ & $135^o$ & $130^o$ & $125^o$ & 1$20^o$ & $115^o$ & $110^o$ \\ \hline \hline
     \end{tabular}
\caption{Table entries are the values we find for $P_{180}$ (the posterior probability for $\delta_{CP}$ to be at $\pi$) using equation~(\ref{eq:p180}), for the prior with $P^{pr}_0=P^{pr}_{180}=0.05$ (or $5\%$). For example if experiments (or global fits) find a gaussian distribution for $\delta_{CP}$ with mean $\mu = 200^o$ and sigma $\sigma = 25^o$, then as per the table the posterior probability is enhanced to $P_{180} = 19\%$ .   Note that for all $\mu$ and $\sigma$ values mentioned in the above table, we find $P_0 \approx 0$.}
\label{tab:r=0.05}
\end{table}


\begin{table}
\begin{tabular}{ l | l | l | l | l | l | l | l | l | l | l | l | l | l | l | l  }
\hline \hline

$\sigma$   \ \ \     $\mu$ & $180^o$ & $185^o$ & $190^o$ & $195^o$ & $200^o$ & $205^o$ & $210^o$ & $215^o$ & $220^o$ & $225^o$ & $230^o$ & $235^o$ & $240^o$ & $245^o$ & $250^o$ \\ \hline \hline
	$35^o$ & 67\% & 67\% & 66\% & 65\% & 63\% & 61\% & 59\% & 55\% & 52\% & 47\% & 43\% & 37\% & 32\% & 27\% & 22\%  \\ \hline
	$30^o$ & 71\% & 70\% & 69\% & 68\% & 66\% & 63\% & 59\% & 55\% & 50 \% & 44\% & 37\% & 31\% & 25\% & 19\% & 14\% \\ \hline
	$25^o$ & 74\% & 74\% & 73\% & 71\% & \cellcolor[HTML]{C0C0C0} 68\% & 63\% & 58\% & 52\% & 44\% & 36\% & 28\% & 20\% & 14\% & 9\% & 5\% \\ \hline
	$20^o$ & 78\% & 78\% & 76\% & 73\% & 69\% & 62\% & 54\% & 44\% & 33\% & 22\% & 14\% & 8\% & 4\% & 2\% & 1\% \\ \hline
	$15^o$ & 83\% & 82\% & 79\% & 74\% & 66\% & 54\% & 39\% & 24\% & 12\% & 5\% & 2\% & 0.6\% & 0.2 \% & 0 & 0 \\ \hline
	$10^o$ & 88\% & 86\% & 81\% & 70\% & 49\% & 24\% & 7\% & 1\% & 0.2 \% & 0  & 0  & 0 & 0 & 0 & 0 \\ \hline
	$5^o$ & 93\% & 90\% & 66\% & 14\% & 0.5\% & 0 & 0 & 0 & 0 & 0 & 0 & 0 & 0 & 0 & 0 \\ \hline 
	$2.5^o$ & 97\% & 79\% & 1\% & 0\% & 0 & 0 & 0 & 0 & 0 & 0 & 0 & 0 & 0 & 0 & 0 \\ \hline \hline
\ \ \ \ \ 	$\mu$ & $180^o$ & $175^o$ & $170^o$ & $165^o$ & $160^o$ & $155^o$ & $150^o$ & $145^o$ & $140^o$ & $135^o$ & $130^o$ & $125^o$ & 1$20^o$ & $115^o$ & $110^o$ \\ \hline \hline
     
\end{tabular}
\caption{$P_{180}$ values with $P^{pr}_0=P^{pr}_{180}=0.25$ (or $25\%$). For example if experiments (or global fits) find a gaussian distribution for $\delta_{CP}$ with mean $\mu = 200^o$ and sigma $\sigma = 25^o$, then as per the table the value of $P_{180} = 68\%$.   Note that for all $\mu$ and $\sigma$ values mentioned in the above table, we find $P_0 \approx 0$.}
\label{tab:r=0.25}
\end{table}
\end{center}
}

\end{widetext}

In Tables~\ref{tab:r=0.05} and~\ref{tab:r=0.25} we display the posterior probability $P_{180}$ evaluated using a prior of $P^{pr}_{0} = P^{pr}_{180}= 0.05$ and $=0.25$ respectively, given that experiments have been fitted by a gaussian distribution for $\delta_{CP}$ with the best fit at $\mu$ (first and last rows of the table) and standard deviation $\sigma$ (first column of table).

For the choices of $\mu$ and $\sigma$ in the table, note that $P_{180} >> P_{0} \approx 0$ as $0^o$ is more standard deviations away from the best-fit $\mu$ than is $180^o$.

Experimental data from $T2K$ and $NOvA$ have been fitted by Nu-Fit 5.0 (2020)~\cite{Esteban:2020cvm} and their current result is $\delta_{CP}=197^{+27}_{-24}$, for the case of Normal Hierarchy of neutrino masses and using data from all neutrino experiments including SK.  The $\mu = 200^o$ and $\sigma =25^o$ corresponding to the shaded regions in Tables~\ref{tab:r=0.05} and~\ref{tab:r=0.25} are the close to the current best-fit results.   We see from the tables that the posterior probability for $\delta_{CP}$ to be at $180^o$ (or within a hundredth of a degree of it) has been enhanced to $19 \%$ (with prior being $5\%$ to be at $180^o$) and to $68\%$ (with 25\% prior to be at $180^o$).   

In next few years the combined sensitivity of  $T2K$ and $NoVA$ may reach $15^o-20^o$ and we see from both the tables that if the best fit point shifts closer towards $180^o$ as more data is taken, then the enhancement of posterior probability $P_{180}$ will be even more (see row corresponding to $\sigma = 15^o$). While if it shifts the other way so that it is more than the current $197^o$ then the enhancement may fall a bit, nevertheless $P_{180}$ remains significant up to $\mu = 230^o$ in Table~\ref{tab:r=0.25}. 

In the next decade new experiments with higher sensitivity such as DUNE and Hyper K will be operational.  The posterior probability $p(\delta_{CP})$ from the  combined fit of current experiments, would then become the prior probability $\pi(\delta_{CP})$ for the Bayseian analysis of the next generation experiments. An alternate way to do a future Bayesian analysis is to use the same priors as we have been using for a global fit of results of all experiments current and future.  

The rows in Tables~\ref{tab:r=0.05} and~\ref{tab:r=0.25} for $\sigma$ corresponding to $15^o$ or less, are relevant for the combined global fit of NoVA, DUNE, T2k and Hyper K.   As we can see from the lower rows of the above tables, if $\delta_{CP} = 180^o$ continues to be within 2-sigma of the experimental best fit point ($\mu$), there will continue to be an enhancement of the posterior probability $P_{180}$ compared to $P^{pr}_{180}$.  And we would need to go beyond a 3-sigma exclusion for $P_{180}$ to drop out of significance.  

On the other hand, if there is a 5-sigma discovery of CP violation in the leptonic sector then of course as can be seen from the tables $P_{180}$ drops to be negligibly small or zero, and our analysis would then become irrelevant.

However our analysis would be of growing importance if $\delta_{CP}$ continues to be within 1-2 sigma (for small sigma, even within 3 sigma) of $180^o$, even as experimental sensitivities increase.  Because in such a situation we cannot accurately determine if $\delta_{CP}$ is at $180^o$ ($\pm 0.01^o$)  or not. In this situation, this Bayesian analysis provides us the next best thing -- that is, the probability $P_{180}$ that $\delta_{CP}$ is at $180^o$ (or within a hundredth of a degree of it).

Before we conclude this section we note that the posterior probabilities in Tables~\ref{tab:r=0.05} and~\ref{tab:r=0.25} correspond to priors $P^{pr}_{0,180}= 0.05-0.25$ that are appropriate for the minimal left-right symmetric model with $P$ and for axionless solutions to the strong CP problem such as Left Right symmetric model with $P$ and $CP$ and the Nelson-Barr model as indicated in Table~\ref{tab:prior}.  

Some experimental groups such as $T2K$ do a Bayesian analysis of their data so as to obtain posterior probabilities~\cite{Abe:2021gky}.  However they only consider a uniform prior for $\delta_{CP}$ (or $sin~\delta_{CP}$) without the delta function like peaks at $0^o$ and/or $180^o$.  Such a prior (with $P^{pr}_{0,180}=0$) is appropriate for the Standard Model augmented by 3 right handed neutrinos, or models based on Grand Unification or for theories that have axions. But it does not suffice for other well motivated ideas such as left-right symmetric model and axionless solutions to the strong CP problem.  

Our analysis shows that it would be worthwhile for experimental groups to also consider Bayesian priors that include delta function like peaks at $0$ and/or $180^o$,  that are appropriate for the axionless solutions to the strong CP problem that we have considered in this work, and for the left-right symmetric model.


\section{Posterior probability using global fit}
\label{sec:postglobal}

\begin{figure}[t]
\includegraphics[width=90mm]{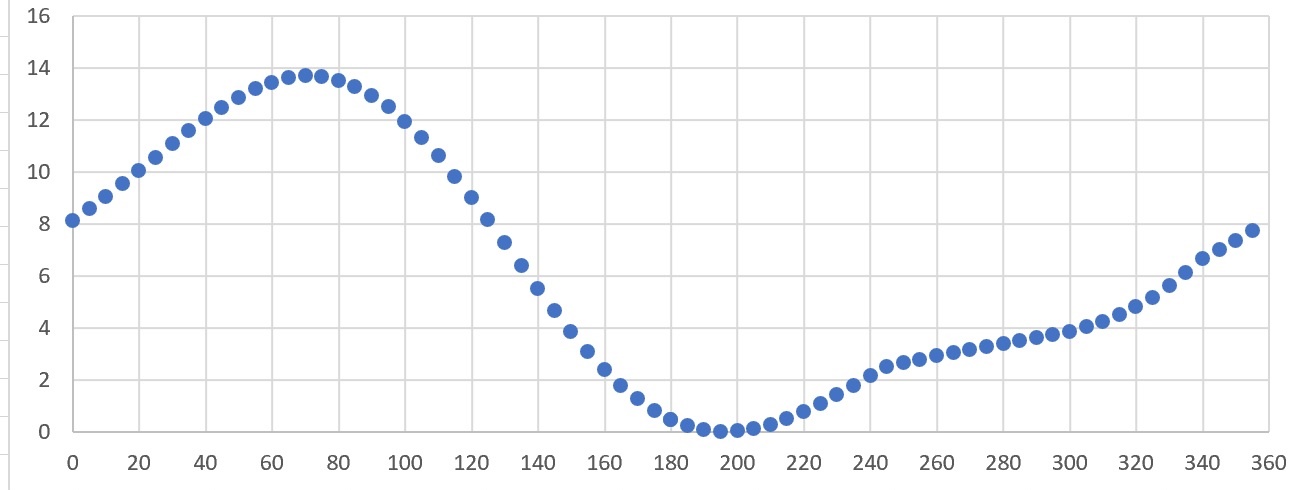}
\caption{The $\chi^2$ distribution for the global fit of current neutrino experimental data from Nu-Fit 5.0 (2020)~\cite{Esteban:2020cvm}}
\label{fig:graph1}
\end{figure}

\begin{figure}[t]
\includegraphics[width=90mm]{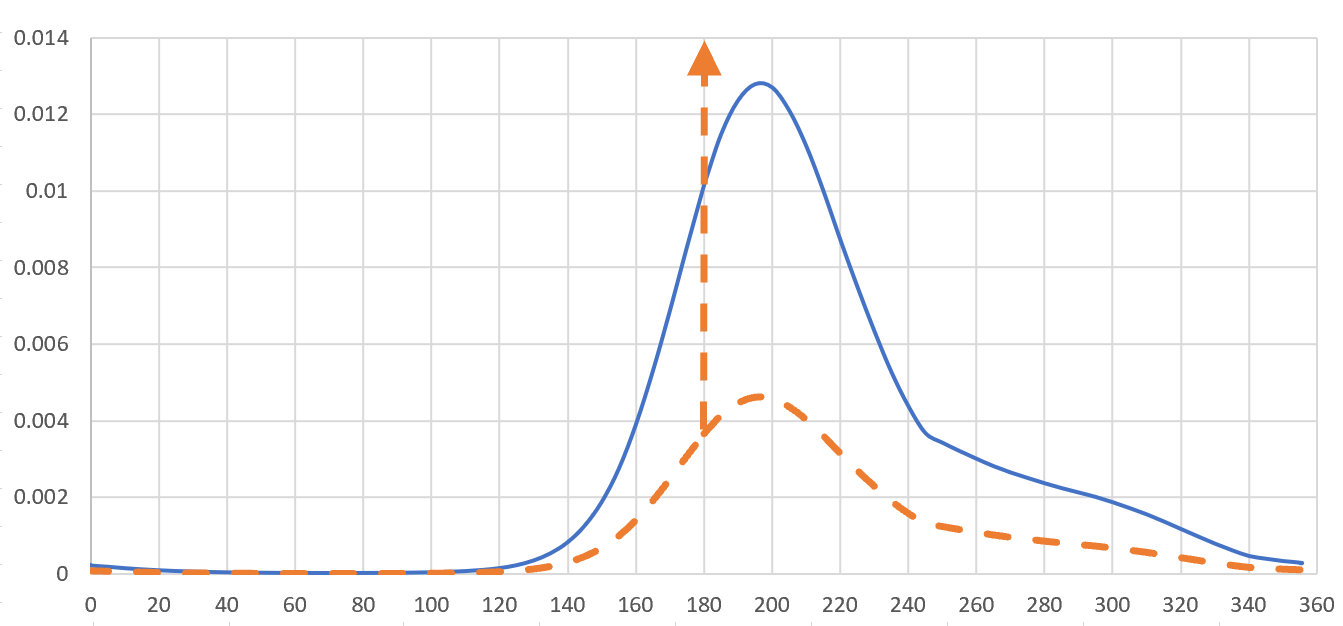}
\caption{The posterior probability density $p(\delta_{CP})$  of equation~\ref{eq:postexpt} (in units of probability/degree), evaluated using the $\chi^2$ from Figure~\ref{fig:graph1} in equation~\ref{eq:like},   for the priors $P^{pr}_{0,180} =0$ (blue or solid curve) and $P^{pr}_{0,180} = 25 \%$ (orange or dashed curve). The area under both the blue (solid) and orange (dashed) curves is 1. In the orange (dashed) curve 64\% of the area is in the delta function at $180^o$. Note that the $P_{180}$ values calculated for the priors used are displayed in Table~\ref{tab:small} and $P_0 \approx 0$. }
\label{fig:graph2}
\end{figure}

We now take $\chi^2$ values from the current global fit for $\delta_{CP}$ and use this to determine the experimental likelihood, using the approximation that log of the likelihood function approximately distributes as $-\chi^2/2$~\cite{wilks1938}.  That is, 

\begin{equation}
L(\delta_{CP}) \propto  e^{-{{\chi^2(\delta_{CP})}\over {2}}}
\label{eq:like}
\end{equation}

where we have taken the $\chi^2$ distribution of the global fit for the single parameter $\delta_{CP}$ by Nu-Fit 5.0 (2020)~\cite{Esteban:2020cvm} (reproduced in Figure~\ref{fig:graph1}) and evaluate the Likelihood distribution using equation~\ref{eq:like}. The blue or solid curve in Figure~\ref{fig:graph2} is in fact proportional to the likelihood distribution.    Note that we have used the Nu-Fit data for normal hierarchy with SK included.

We now use equation~(\ref{eq:posteq}) along with the likelihood distribution obtained from the $\chi^2$ of Fig~\ref{fig:graph1} to evaluate the posterior probabilty for priors with $P^{pr}_{0}= P^{pr}_{180} = 0.05, 0.25$ in equation~(\ref{eq:prioreq}).   

				\begin{table}[t]
\begin{tabular}{  l | l  l  l  l }
\hline \hline

$P^{pr}_{0,180}$  \   &    $P_{180}$ \ \ & $P^{\pm 1}_{180}$ \ \ & $P^{\pm 5}_{180}$ \ \ & $P^{\pm 10}_{180}$ \\ \hline \hline
0\%  & 0 & 2\% & 10\% & 18\% \\ \hline
5\%	 & 17\% & 19\% & 25\% & 32\% \\ \hline
25\% & 64\% & 64\% & 67\% & 70\% \\
	 \hline \hline
     
\end{tabular}
\caption{Using the global fit to $\delta_{CP}$ from the current results of neutrino experiments,  the table displays $P_{180}$, the posterior probability to be within a hundredth of a degree of $180^o$ and $P^{\pm n}_{180}$, the posterior probability to be within $n$ degrees of $180^o$, for different values of the priors $P^{pr}_{0,180}$.} 
\label{tab:small}

\end{table}

The posterior probability distribution can be expressed in the form 

\begin{eqnarray}
\label{eq:postexpt} 
p(\delta_{CP}) & = & P_0 \delta(\delta_{CP}) + P_{180} \delta(\delta_{CP}-\pi) \nonumber \\  & \  & + \left(1-(P_0+P_{180})\right) L'(\delta_{CP})
 \end{eqnarray}
where $L'(\delta_{CP}) = L(\delta_{CP})/N'$ is normalized using $N'= \int^{2\pi}_0 L(\delta_{CP}) d\delta_{CP}$.

We can now evaluate $P_{0,180}$ using eqns~(\ref{eq:normN}) and~(\ref{eq:p180}) with $g$ replaced by $L'$ and $\mu, \sigma$ dropped.  We find $P_0 \lesssim 1\% \approx 0$ and
\begin{eqnarray}
P_{180} & = & \ 0\% ~for~P^{pr}_{0,180}=0 \% \nonumber \\
& = & 17\% ~for~P^{pr}_{0,180}=5 \%  \nonumber \\
        & = & 64 \% ~for~P^{pr}_{0,180}=25 \% 
				\label{eq:compare}
				\end{eqnarray}
				
								The posterior probability density distribution $p(\delta_{CP})$ that we obtain, has been plotted in Figure~\ref{fig:graph2}.

				For $P^{pr}_{0,180}=5 \%$ and $25\%$, we can compare the values obtained in eqn~\ref{eq:compare}  with the $P_{180}$ values in the shaded regions of Tables~\ref{tab:r=0.05} and \ref{tab:r=0.25} and see that they are close and the Gaussian approximation of the previous section works remarkably well.

				In Table~\ref{tab:small} we display  $P^{\pm n}_{180}$,  the probability that $\delta_{CP}$ is small, that is within $n^o$ of $180^o$.  We evaluate $P^{\pm n}_{180}$ by integrating the posterior probability distribution of eqn~(\ref{eq:postexpt}) using $L'$.  
				
				\section{Utility of posterior probability}
				\label{sec:utility}
				
				To understand the utility of the posterior probability $P_{180}$ let us recall the next to the minimal version (with one heavy lepton family added) of the axionless left-right symmetric strong CP solution with $P \times CP$ of Section~\ref{sec:SCPLR}.  In this model whether tree level leptonic CP phases are generated or not depends on whether the intrinsic parity of the heavy leptons that can have CP violating mixings with the usual leptons is $1$ or $i$. If it is $1$ like the usual leptons, then leptonic CP phases are generated, if it is $i$ then then $\delta_{CP}$ would be be either $0$ or $\pi$ at the tree level.  A priori we would give a 50-50 chance for the intrinsic parity to be $1$ or $i$.  That is we'd take $P^{pr}_{0,180} = 0.25$ as argued in section~\ref{sec:SCPLR}. 
				
				As can be seen from Table~\ref{tab:small}, the posterior probability $P_{180}$ given the current experimental likelihood for this prior is $64\%$... which basically means that the posterioir probability of the intrinsic parity being $i$ has gone up from the prior of $50\%$ to $64\%$, for this model.  As the experimental sensitivity increases if the best fit value of $\delta_{CP}$ moves closer to $180^o$ then as shown in Table~\ref{tab:r=0.25} $P_{180}$ can become higher than $90\%$.  That is if nature adopted this model, we would be able to figure out with greater and great confidence that the tree-level CP phase is real or equivalently, that the intrinsic parity of the heavy lepton is $i$. 
				
				While the above is how the analysis is useful for specific models, it also has a more general utility.  Regardless of what the true theory of nature is if we allow for some probability that it may be a theory that doesnt have axions and that some such theories can have large fractions of their parameter space where CP is conserved in the leptonic sector (like it is for the strong CP phase) to within a fraction of a degree or less, and we assign a low prior probability such as $10\%$ for being in this parameter space, while our prior for there being leptonic CP violation is $90\%$ distributed uniformly from $0$ to $2\pi$ due to a belief in axions or the standard model or GUTs, then an enhanced value for the posterior $P_{180}$ from Table~\ref{tab:r=0.05} will reduce our belief in the paradigm of SM-axions-GUTs and increase our belief in alternate ideas such as such as minimal left-right model, or axionless strong CP solutions involving $CP$ conservation such as $P\times CP$ or Nelson-Barr.
				
				On the other hand if $P_{180}$ gets diminished to less than $1\%$ as experimental sensitivity and data increase, then we would get the measurement of the leptonic CP phase and a potential 5-sigma discovery of leptonic $CP$ violation. 
				Amongst the models we discussed,  the one that would be ruled out with discovery of leptonic CP violation is the minimal version of the axionless strong CP solving left-right model with $P \times CP$ described initially in section~\ref{sec:SCPLR} without the inclusion of a heavy vector-like lepton family.  
				

\section{Conclusions}
\label{sec:conc}

In this work we obtained the posterior probability distribution for the leptonic CP phase $\delta_{CP}$ by using a prior distribution that has delta function like peaks at $0$ and $\pi$ on top of a uniform distribution.  Such a prior allows us to compare different theoretical expectations for $\delta_{CP}$ based on the choice of solutions to the strong CP problem, or its implications in models where it is not solved.  

We find that the posterior probability for $\delta_{CP}$ to be at $180^o$ (or within a percent of degree of it) is enhanced to $P_{180}=64\%$ (from prior probability $P^{pr}_{180}=25\%$) when it is evaluated using Nu-Fit's Global fit to current data (with Normal Hierarchy) from neutrino experiments including $T2K$ and $NOvA$.  

We also provide helpful tables of the posterior probability for potential results from future experiments at higher sensitivity.

\section*{Acknowledgment}  
Thanks to Sonali Tamhankar for illuminating discussions that led to this work.

\bibliography{LRsymmetrythi}

\end{document}